\documentclass[12pt]{iopart}
\usepackage{braket}
\usepackage{amsfonts}
\usepackage{graphicx}
\usepackage{hyperref}
\usepackage{caption}

\begin{document}

\title[Quantum Gravity at the Fifth Root of Unity]{Quantum Gravity at the Fifth Root of Unity}

\author{Marcelo Amaral, Raymond Aschheim, Klee Irwin}

\address{Quantum Gravity Research\\
				{\small \em Los Angeles, CA}}

\ead{Marcelo@QuantumGravityResearch.org, Raymond@QuantumGravityResearch.org and Klee@QuantumGravityResearch.org}

\vspace{10pt}
\begin{indented}
\item[]\today
\end{indented}

\begin{abstract}

We consider quantum transition amplitudes, partition functions and 
observables for 3D spin foam models within $SU(2)$ quantum group deformation symmetry,
where the deformation parameter is a complex fifth root of unity. By considering
``fermionic'' cycles through the foam  we couple this $SU(2)$ quantum group
with the same deformation of $SU(3)$, so that we have quantum numbers linked with
spacetime symmetry and charge gauge symmetry in the computation of observables. 
The generalization to higher-dimensional
Lie groups $SU(N)$, $G_2$ and $E_8$ is suggested. 
On this basis we discuss a unifying framework for quantum gravity. 
Inside the transition amplitude or partition function for geometries, we
have the quantum numbers of particles and fields interacting in the form of a spin foam network
$-$ in the framework of state sum models, we have a sum over quantum computations driven by the 
interplay between aperiodic order and topological order.

\end{abstract}

\vspace{2pc}
\noindent{\it Keywords}: Quantum Gravity, Spin Foam, Unification Physics, Aperiodic Order, Topological Order


\section{Introduction}
\label{intro}

Quantum gravity and unification physics programs, in the absence of more concrete experimental results, rely much
on rigorous mathematical results to make progress. The old and well established quantum gravity programs of
 loop quantum gravity (LQG) \cite{Rovelli-Vidotto-Book} and string theory \cite{stringbook1,Schwarz} are good examples of this new paradigm
to find clues about the underlying physics at the Planck scale.
For a review of the motivations and main results of LQG and spin foam models we recommend
the book \cite{Rovelli-Vidotto-Book}. We also recommend a review of the main elements of the higher dimensional Lie algebra
unification program  \cite{Georgi,Koca3,Koca4,Zee,gilmore}, which is closely related to string theory.
With those results as guides, this paper will construct a model 
that does not have a priori assumptions of general relativity (GR) and the usual Yang Mills gauge quantum fields,
but yet has their building blocks built-in in a consistent way to generate their emergent macroscopic properties
starting only from the notion of quantum geometry.

It is interesting to point out that LQG starts from GR and provides its quantization which leads to the 
quantization of the geometry itself; lengths, areas and volumes, in operator form, have discrete spectra; they have minimum quanta.
But it does not incorporate matter and the quantum fields $-$ the charge space symmetries.
On the other side, string theory starts from the quantum fields and implements the generalization 
from point particles to extended particles,
like a one dimensional string or the higher dimensional branes. This also leads to discrete spectra, and a more complete
description of unification physics. But this unification picture is more complicated because
the strings are defined over a spacetime manifold, which leads to fundamental problems and the search
for a more fundamental $(M-)$theory, where one hopes to understand the existence of a spacetime manifold itself 
as the emergent property of a specific vacuum rather than an
identifiable feature of the underlying theory \cite{Schwarz}. A partner approach of string theory,
but less ambitious, is the Lie algebraic unification program, as in $E_8$ unification \cite{LisiLieE8,E8CarlosonTony}.
This makes use of the generalization of the simplest non-abelian gauge symmetry, $SU(2)$, to higher-dimensional ones, ending with
a complete unification of the charge Lie group symmetries of the standard model in the largest
exceptional Lie algebra $E_8$, making heavy use of the representation theory of these algebras.
Our aim is to use this unification aspect of the representation theory of Lie algebras to improve
the spin foam approach of LQG. We can say that spin foam partition functions or
transition amplitudes are about the interaction of representations of spacetime symmetry, and so, ultimately,
$E_8$ can provide the network of interaction representations incorporating the usual charge symmetries $-$ but
here we focus on the first step in 3 dimensions where spacetime symmetry is just $SU(2)$ and internal charge
symmetry is $SU(3)$.

Both the LQG and string/M-theory programs lead to the idea of building blocks for all the fields that constitute our
classical and quantum worlds, from which the
question arises of how to ``glue'' these fundamental blocks together in an elementary and first-principles way. 
There have been 
recent efforts to use the idea of entanglement in both approaches \cite{EugenioNelson,Raamsdonk}. 
Our recent work \cite{KleeCodetheoreticaxiom,5throot} pointed out that there is a 
principle correlated with the physical realization
of representations of Lie groups and algebras.
This indicates, in accordance with the results of LQG and string theory, 
that quantization of spacetime must respect a special kind
of quantum symmetry, implemented as a code made of a small set of representations
of its symmetry group, each with specific fusion  rules and syntactical degrees of freedom 
that express physical ``observables'' in the form of quantum fields at large scales.
This concept is in line with quantum information and digital physics principles applied to 
spacetime, and aims to bring new advances in quantum gravity and unification physics.

In the model we develop here the building blocks are 
very constrained. There is a small set of Lie algebra representations that interact, and the syntax of its
fusion rules works as 
a glue, indicating topological order \cite{5throot,ZhenghanWang,Pachos}. 
With a tetrahedron building block the problem is translated to an issue of 
how to tile the space, where we will be mainly concerned with 3-dimensional space,
which can be linked with the Fibonacci anyonic fusion Hilbert space $-$ a computational space.
The problem of tiling the space brings us to aperiodic order, which generalizes
the notion of long range periodic order $-$ from lattices to quasicrystals \cite{Senechal,Fang-Irwin}.
Quasicrystals also play an important role in aspects of the representation theory of Lie algebras 
\cite{ChenMoodyPatera,Koca1,Koca2} that are important
for the connection with spin foam models.

This paper is organized as follows: in Section~\ref{sec:spinfoam} we review the usual concept of 
3D spin foam quantum gravity, and present its fifth root of unity ``quantization'' as well as
the Fibonacci fusion Hilbert space interpretation. In Section~\ref{sec:unification} we discuss
the coupling of the model with internal $SU(3)$ symmetries and suggest a
natural extension to the larger $SU(N)$, $G_2$, and $E_8$, with the same fifth root of unity
deformation, and compute some observables. We present our conclusions in Section~\ref{sec:conclusion}.

\section{Spin Foam Code or Sum Over Quantum Computations}
\label{sec:spinfoam}

Following \cite{Rovelli-Vidotto-Book}, mainly chapters 5 and 6,
we consider a discretization of spacetime in terms of a triangulation ($\triangle$) and its dual 2-complex ($\triangle$*).
A three-dimensional triangulation is given by tetrahedrons, triangles, segments and points.
For a four-dimensional triangulation, we include the 4-simplex.
The dual 2-complex in three-dimensional (3D) bulk is made by associating a tetrahedron with a vertex, 
a triangle with an edge, and a segment with a face. On the boundary the dual of triangles are nodes and segments are links.
In four dimensions (4D), the dual of a 4-simplex is a vertex, the dual of a tetrahedron is an edge and the dual of a triangle
is a face. 
The boundary graph and terminology are equal to 3D.
The gauge group of symmetry is $SU(2)$ in 3D, the covering group  of the rotation group $SO(3)$, and $SL(2,C)$ or
$SU(2)\times SU(2)$ in 4d,
the respective covering groups of the Lorentz group and the 4D rotation group $SO(4)$. 
The variables are $SU(2)$ group elements $U_e$ associated to edges $e$ in the bulk of $\triangle$* or  $U_l$ associated to links 
$l$ on the boundary, and algebra elements $L_f=L^i_fJ_i$ 
associated to faces $f$ of $\triangle$* or $L_l=L^i_l J_i$ associated to links on the boundary. $J_i$ are the $SU(2)$ generators,
$J_i=\frac{i}{2}\sigma_{i}$, where $\sigma_{i}$ are the usual Pauli matrices.
The quantization involves operators $U_l$ and $L_l$ on the boundary realizing a commutation relation on a Hilbert space,
\begin{equation}
\left[U_{l},L_{l'}^{i}\right]=i\beta\delta_{ll'}U_{l}J^{i}\label{eq:commutationUL}
\end{equation}
and
\begin{equation}
\left[L_{l}^{i},L_{l'}^{j}\right]=i\beta\delta_{ll'}\varepsilon_{k}^{ij}L_{l}^{k},\label{eq:commutationsu2algebra}
\end{equation}
where $\beta$ is a constant related to the gravitational constant, $\delta_{ll'}$ is the Kronecker-delta
and $\varepsilon_{k}^{ij}$ is the totally antisymmetric Levi-Civita symbol.
States are wavefunctions $\Psi(U_{l})$ or $\Psi(L_{l})$ of $L$ group/algebra elements on $L$ links of the boundary graph 
modulo the $SU(2)$ gauge symmetry implemented on the nodes. The Hilbert space is the space of square integrable
functions of these coordinates:
\begin{equation}
\mathcal{H}_{\Gamma}=L_{2}\left[SU(2)^{L}/SU(2)^{N}\right]_{\Gamma}\label{eq:hilbertspaceLQG}
\end{equation}
where $\Gamma$ is the boundary graph.
The gauge invariant states must satisfy $\Psi(U_{l})=\Psi\left(\lambda_{s_{l}}U_{l}\lambda_{t_{l}}^{-1}\right)$ 
for every node $n$ of the boundary graph.
One major realization of loop quantum gravity is that the kinematics are the same in 3D and 4D $-$ the Hilbert space and commutation
relations are in both cases the ones for the $SU(2)$ group. The distinction from 3D to 4D arises only in the dynamics, 
with the 4D gauge symmetry  $SL(2,C)$ or $SU(2)\times SU(2)$  implemented on the bulk of the spin foam.

The dynamics are implemented in the form of a state sum model with quantum transition amplitudes or partition functions.
Let us consider first the transition amplitude. It is a function of the states defined on $\Gamma=(\partial\triangle)^*$,
the boundary graph. This can be defined in the group representation $W_{\triangle}(U_l)$ or in the so called 
spin representation $W_{\triangle}(j_f)$, where the $\triangle$ indicates that the amplitude is computed
in a discretization $\triangle$ of the bulk that matches the boundary graph. 
In $W_{\triangle}(j_f)$, a basis on the gauge invariant Hilbert space is given by the normalized 
eigenvectors of the operator $L_l$, indicated by $\ket{j_f}$. An element of this basis is determined by assigning a spin $j_l$
of a representation of $SU(2)$ to each link $l$ of the graph. A graph with a spin assigned to each link is called a spin network.
So the spin network states $\ket{j_f}$ form a basis of the gauge invariant Hilbert space of quantum gravity, i.e. they span
the quantum states of the geometry. The quantum transition amplitude is given by
\begin{equation}
W_{\triangle}(j_{l})=\mathcal{N}_{\triangle}\sum_{j_{f}}\prod_{f}A_{j_{f}}\prod_{v}A_{v}(j_{f}),\label{eq:transitionamplitudegeneral}
\end{equation}
where $\mathcal{N}_{\triangle}$ is a normalization constant that can depend on the discretization, and there is an amplitude $A_{j_f}$
for each face of $\triangle$* and an amplitude $A_v(j_f)$ for each vertex, both in 3D and 4D. 
Equivalently, in 3D, the amplitude $A_{j_f}$ can be defined
on the segments of $\triangle$ and $A_v(j_f)$ on the tetrahedrons\footnote{The input for $A_v(j_f)$ are the representations on
each one of the 6 faces adjacent to the vertex or the representations on the 6 segments that form the tetrahedron. We can note 
that $A_v(j_f)$ comes from a path integral quantization and gives the weighted value of the field at one point of the
discretization.}. Similarly, in 4D the amplitude $A_{j_f}$ can be defined
on the triangles of $\triangle$ and $A_v(j_f)$ on the 4-simplexes\footnote{The 4-simplex has 10 edges and 10 triangles,
so the input for $A_v(j_f)$ are 10 representations. The boundary graph of the amplitude is a pentagram.}. 
The partition function ($Z_{\triangle}$) in the situation with the triangulation without boundaries is the same
as eq.~(\ref{eq:transitionamplitudegeneral}) but summing also over the boundary representations.

From now on we consider the 3-dimensional spin foam model, but it was necessary
 to point out in the preceding review, that 
the step from 3D to 4D is not a big one.
In the spin network basis we have that the amplitudes $A_{j_{f}}$
are given by the dimension of the representation, $d_{j}$, and the
vertex amplitude $A_{v}(j_{f})$, which implements the $SU(2)$ symmetry,
is a function that takes the 6 spin quantum numbers of the representations
on the 6 edges of the tetrahedron around a vertex and returns a complex
number. The vertex amplitude in this case is given mainly by the Wigner 6j-symbol 
 $\{6j\}=\left\{ \begin{array}{ccc}
j_{1} & j_{2} & j_{3}\\
j_{4} & j_{5} & j_{6}
\end{array}\right\} $\footnote{See \cite{Rovelli-Vidotto-Book} for explicit definitions.}, and the transition
amplitude is
\begin{equation}
W_{\triangle}(j_{l})=\mathcal{N}_{\triangle}\sum_{j_{f}}\prod_{f}(-1)^{j_{f}}d(j_{f})\prod_{v}(-1)^{j_{v}}\{6j\},\label{eq:transitionamplitude3d6j}
\end{equation}
where $j_{v}=\sum_{a=1}^{6}j_{a}$.

\begin{figure}[!h]
	\centering{}
	\includegraphics[scale=0.30]{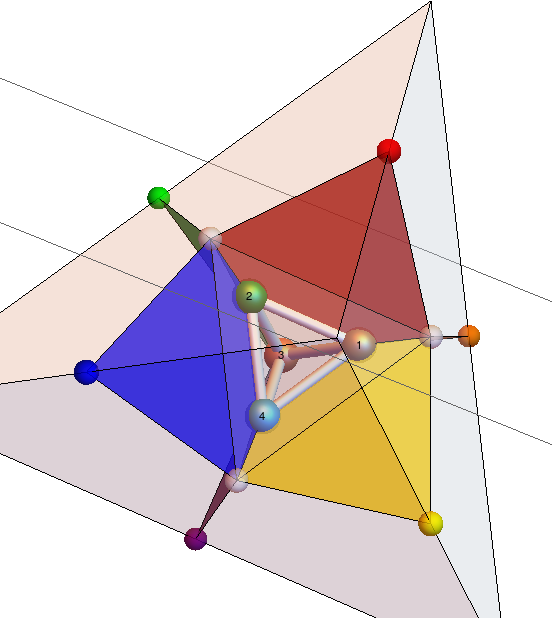}
	\caption{The dual 2-complex for the triangulation level 1 with four tetrahedrons, highlighting the external faces.}
	\label{triangduallevel1}
\end{figure}

Note that even with the definition of this object on a truncation of the triangulation, still there are non-physical divergences.
We see this by first writing explicitly eq.~(\ref{eq:transitionamplitude3d6j}) for a triangulation made of 1 tetrahedron. The dual
is made of a vertex inside the tetrahedron connected to a node on each of the 4 triangular faces.  The dual has 6 boundary faces that we label with 6 $SU(2)$ representations (or equivalently there is a dual tetrahedron and the representations are living on its edges) and the transition amplitude is immediately 
\begin{equation}
W_{\triangle_{1}}(j_{l})=(-1)^{j_{v}}\left\{ \begin{array}{ccc}
j_{1} & j_{2} & j_{3}\\
j_{4} & j_{5} & j_{6}
\end{array}\right\} \label{eq:transitionamplitudedelta1}
\end{equation}
and is well defined, although for the partition function one should sum
over the infinite number of $SU(2)$, $j_{f}$, representations. The problem
for transition amplitudes occurs already for the triangulation with
only four tetrahedrons, made by inserting a point inside the initial
large one and connecting it with the 4 tetrahedron points. The dual
is made of four vertices inside the 4 tetrahedrons connected with each
other and the boundary faces. Now there are the 6 boundary faces with
representations labeled $j_{1}...j_{6}$, plus 4 internal faces, $j_{7}...j_{10}$,
of the dual, (see Figure~\ref{triangduallevel1}). 
What happens is that even with the 6 boundary spin representations
fixed, the classical $SU(2)$ fusion rules don't give a constraint on
the spin representations on these internal faces, and the amplitude
can diverge
\begin{equation}
W_{\triangle_{4}}(j_{1},...,j_{6})=\sum_{j_{f}}\prod_{f=1}^{10}(-1)^{j_{f}}d(j_{f})\prod_{v=1}^{4}(-1)^{j_{v}}\{6j\},\label{eq:transitionamplitude4tets}
\end{equation}
where each $\{6j\}$ now has 3 boundary spin representations and 3
internal ones. These internal spins are not fixed by triangular inequalities. 
All the four internal spin form an internal
tetrahedron where the spins can flow without restriction $-$ the tetrahedron with the vertices 1, 2, 3 and 4 in 
Figure~\ref{triangduallevel1}. From results of LQG, the spin quantum numbers correspond
to eigenvalues of geometrical operators such as lengths and areas, and so
there is a divergence when the scale of the geometry becomes large, a kind of infrared divergence.
This is possible because in Riemannian geometry the space can be strongly
curved. The
4 internal faces form what is called a bubble in LQG, and the divergence
for a large spin sitting on an internal triangular face is called
a spike. 
 
The transition amplitudes can be made well defined and explicitly
independent of the triangulation by a deformation of the $SU(2)$
symmetry, using quantum $SU(2)_q$ with a deformation parameter $q$ defined
as a complex root of unity $q=e^{\frac{2\pi i}{r}}$, with $r$ an
integer\footnote{See chapter 6 of \cite{Rovelli-Vidotto-Book}.}.
The transition amplitudes is given by 
\begin{equation}
W_{q}(j_{l})=\mathcal{N}_{q}\sum_{j_{f}}\prod_{f}(-1)^{j_{f}}d_{q}(j_{f})\prod_{v}(-1)^{j_{v}}\{6j\}_{q},\label{eq:transitionamplitudequantumgroup}
\end{equation}
where we have now, as the quantum-deformed analogues of the objects
that appear in the transition amplitude, the quantum dimension
$d_{q}(j_{f})$ and the quantum Wigner 6j-symbol $\{6j\}_{q}=\left\{ \begin{array}{ccc}
j_{1} & j_{2} & j_{3}\\
j_{4} & j_{5} & j_{6}
\end{array}\right\} _{q}$. From the representation theory of quantum groups \cite{Biedenharn}
we know that the spin quantum numbers of $SU(2)_{q}$ are constrained
to $j\leq\frac{r-2}{2}$, working as a cut off for the flow of spin
on the bubbles. 
The usual triangular inequalities are modified and
supplemented by the conditions  $2j_{1},2j_{2},2j_{3}\leq j_{1}+j_{2}+j_{3}\leq r-2$,
which are the triangular inequalities of a triangle
on a sphere with a radius determined by $r$. The geometry of $q$-deformed
spin networks is therefore consistent with the geometry of constant
curvature space, with the curvature determined by the specific root of unity.

Our particular synthesis of this result is that a general
large spin transition amplitude should be built from ``gluing''
quantum building blocks $W_{\triangle_{1}}$ given by $\{6j\}_{q}$
with a specific $q$ that fixes $W_{\triangle_{1}}$ to low spin representations,
and we will justify in the following the choice $q=e^{\frac{2\pi}{5}i}$.

\subsection{The Quantum Tetrahedron with Topological Symmetry}
\label{qt}

We propose to derive the large spin transition amplitude previously discussed from
the direct quantization of geometry, in particular a quantum building block, 
 a tetrahedron, following the notion of quantum tetrahedrons \cite{Barbieri}.
The quantization of one tetrahedron allows us to define and address quantization 
of geometric quantities like lengths, areas and volumes. Yet the usual quantization with 
$SU(2)$ symmetry
is much too general, allowing an infinity of values for these geometric quantities.
We therefore follow the conceptual discussion in \cite{KleeCodetheoreticaxiom},
which states that the Planck scale regime is a code $-$ \textit{(1) a finite set of symbolic objects, with
(2) ordering rules and (3) syntactical freedom (4) for the purpose of expressing meaning, e.g.,
self-referential physical observables and expected values}. More specifically, the syntactical
rules of the code are related 
to restrictions on the representation theory of Lie algebras \cite{5throot}, which can
be implemented by the usual deformation of the classical algebra with a complex root of 
unity parameter.

Let us discuss a specific model implementation. Most of the geometric
quantities of interest such as lengths, areas and volumes can be described
in a unifying way and simplest form from the geometry of the 3-simplex,
the tetrahedron. The tetrahedron can be described by its 4 (outgoing)
normals, vectors $\mathbf{L}_{a}$ in its four triangular faces, $a=1,2,3,4$,
subject to the closure constraint 
\begin{equation}
\mathbf{C}=\sum_{a=1}^{4}\mathbf{L}_{a}=0.\label{eq:classicalclosureconstraint}
\end{equation}
All geometric properties of the tetrahedron can be derived from the
normals and must be invariant under a common $SO(3)$ rotation of
the tetrahedron. For example, the area of face $a$ is given by $A_{a}=|\mathbf{L}_{a}|$
and the volume is defined by $V^{2}=\frac{2}{9}\left(\mathbf{L}_{1}\times\mathbf{L}_{2}\right)\cdot\mathbf{L}_{3}$.
The proposed quantizing postulate is a quantum deformation of eq.~(\ref{eq:commutationsu2algebra}).
Eq.~(\ref{eq:commutationsu2algebra}) is the $SU(2)$ algebra implied by the 3 dimensional rotational
symmetry, where the normals are promoted to operators $L_{a}$ and
identified with the generators of the algebra $L_{a}^{i}$ ($i=1,2,3$).
We first write the classical algebra in terms of Cartan generators and root vectors, which are
the usual ladder operators of $SU(2)$, then choose
the Cartan generator $h_{a}=L_{a}^{3}$ and the root vectors $L_{a}^{\pm\alpha}=L_{a}^{1}\pm\alpha iL_{a}^{2}$,
where for $SU(2)$, $\alpha=1$, and we have 
\begin{eqnarray}
\left[h_{a},L_{b}^{\pm}\right] & =\pm\beta\delta_{ab}L_{a}^{\pm}\nonumber \\
\left[L_{a}^{+},L_{b}^{-}\right] & =2\beta\delta_{ab}h_{a},\label{eq:commutationJplusminos-1}
\end{eqnarray}
where $\beta$ has the dimension of area. The usual Casimir invariant
is $L_{a}^{i}L_{a}^{i}=L_{a}^{+}L_{a}^{-}+h_{a}\left(h_{a}+\beta\right)$,
commuting with each element of the algebra, and has eigenvalues\footnote{Splitting
the constant with dimension of area $j_{a}\rightarrow\beta j_a$.} ${\beta^{2}j_{a}(j_{a}+1)}$,
$j_{a}\in{0,\frac{1}{2},1,\frac{3}{2},2,...}$, which gives a quantization
of area $A_{a}=\beta\sqrt{j_{a}(j_{a}+1)}$ for a triangular face
or $A_{j_{an}}=\beta\sum_{n}\sqrt{j_{an}(j_{an}+1)}$ for an arbitrary
surface punctured by N punctures $n_{1},...,N$. As discussed in the 
previous section, there are no restrictions on the flow of spin representations in 
this case, suggesting we should implement a more fundamental topological symmetry.
So we apply the $SU(2)_q$
deformation at the fifth root of unity  $q=e^{\frac{2\pi i}{r}}$,
with $r=5$ (from now on let us call it $SU(2)^5_q$), which affects the Cartan generator in the second term
and constitutes essentially a quantization postulate for geometry:
\begin{eqnarray}
\left[h_{a},L_{b}^{\pm}\right] & =\pm\delta_{ab}L_{a}^{\pm}\nonumber \\
\left[L_{a}^{+},L_{b}^{-}\right] & =2\delta_{ab}\left[2h_{a}\right]_{q},\label{eq:commutationJplusminos-1-1}
\end{eqnarray}
with the q number $\left[x\right]_{q}$ defined by 
\begin{equation}
\left[x\right]_{q}=\frac{q^{\frac{x}{2}}-q^{-\frac{x}{2}}}{q^{\frac{1}{2}}-q^{-\frac{1}{2}}}=\frac{sin\left(\frac{\pi}{5}x\right)}{sin\left(\frac{\pi}{5}\right)}\label{eq:qnumber}
\end{equation}
or writing the Cartan generator, $\mathcal{J}_{a}=q^{h_{a}}$: 
\begin{eqnarray}
\mathcal{J}_{a}L_{b}^{\pm}\mathcal{J}_{a}^{-1} & =q^{\pm1}L_{b}^{\pm}\nonumber \\
\left[L_{a}^{+},L_{b}^{-}\right] & =2\delta_{ab}\frac{\mathcal{J}_{a}-\mathcal{J}_{a}^{-1}}{q-q^{-1}}\label{eq:hopfalgebra}
\end{eqnarray}
For this deformation to be dimensionally consistent, we start with everything
adimensional and there is no need a priori for $\beta$. There is
a notion of scale incorporated in the deformation parameter, $r=\frac{l_{c}}{l_{p}}$,
with $r$ an integer and the large scale $l_{c}$ given in units of the
underlying scale $l_{p}$. This representation
theory is as well understood as that of $SU(2)$. There are 3 independent
Casimir invariants for each face $C_{aq}=L_{a}^{+}L_{a}^{-}+\left[h_{a}\right]_{q}\left[h_{a}+1\right]_{q}$,
$X_{a}=(L_{a}^{+})^{5}$ and $Y_{a}=(L_{a}^{-})^{5}$. When $X_{a}$
and $Y_{a}$ have zero eigenvalues for all vectors in the linear vector
space carrying the irreducible representation, the representations
have classical analogs, and can be constructed by specializing from
the general values of $q$ to the specific root of unity. 
In this case the spin quantum numbers that label the representations
of $SU(2)_{q}$ are constrained to $j_{a}\leq\frac{3}{2}$, and are
related to the Casimir $C_{aq}$ of eigenvalues $[j_{a}]_{q}[(j_{a}+1)]_{q}$.
In this way the area spectrum interpretation 
\begin{equation}
A_{a}=\sqrt{[j_{a}]_{q}[(j_{a}+1)]_{q}}\label{eq:quantumarea}
\end{equation}
is not immediate and a scale should emerge in an appropriate limit.

The vertex amplitude is naturally given by the specialization of eq.~(\ref{eq:transitionamplitudedelta1})
to $SU(2)^5_q$ symmetry defined by the Hilbert space of the tensor
product of 4 representations at the vertex $H_{j_{q1}}\times H_{j_{q2}}\times H_{j_{q3}}\times H_{j_{q4}}$,
and it is a function of the representations on the 6 boundary faces
defined by this vertex 
\begin{equation}
W_{\triangle_{1}}^{5}(j_{lv})=(-1)^{j_{v}}\left\{ \begin{array}{ccc}
j_{1} & j_{2} & j_{3}\\
j_{4} & j_{5} & j_{6}
\end{array}\right\} _{5},\label{eq:transitionamplitudedelta1-5}
\end{equation}
with $j_{lv}=j_{1},...,j_{6}$ and the index 5 representing the $SU(2)^5_q$
deformation. The q-deformed 6-j symbols, $\left\{ 6j\right\} _{5}$,
for the 4096 combinations of $j$'s in ${0,1/2,1,3/2}$, are all null
except for a small set of combinations, which can be computed from
eq.(\ref{eq:symbol}) restricted to the $SU(2)$ case and takes on only seven
values: $\{0,\pm1,\pm\varphi,\pm\sqrt{\varphi}\}$, where $\varphi=\frac{\sqrt{5}-1}{2}=\phi^{-1}$,
with $\phi=\frac{\sqrt{5}+1}{2}=1.618...$, the golden ratio.

More generally, this kind of amplitude is independent of the triangulation, 
and gives a topological invariant \cite{Turaev92}, which
means that it does not depend on the triangulation inside the 3 manifold,
but depends only on the edge values and on the topology of
the triangulated 3 manifold (the number of punctures on the boundary 2-sphere). 
Exploiting the possibility to build invariants, by fixing the representations
labels on the edges, we can consider networks of these
topological amplitudes, which act as building blocks for decomposing a higher dimensional
tensor product space, which we will motivate more in the next
section. For example, we can consider amplitudes, functions of representations
on boundary faces of edges for closed loops along the 3 dimensional spin foam
\begin{equation}
W_{c}(j_{lc})=\mathcal{N}_{q}\sum_{j_{f}|j_{lc}}\prod_{f}(-1)^{j_{f}}d_{q}(j_{f})\prod_{v}W_{\triangle_{1}}^{5}(j_{lv}),\label{eq:transitionamplitudequantumgroup-5}
\end{equation}
where $c$ is a closed loop across the 3 dimensional triangulation
and $j_{lc}=j_{1},...,j_{n}$ are the representations on the boundary faces
of the $n$ edges contained in $c$. The sum $\sum_{j_{f}|j_{lc}}$ is taken while maintaining those
boundary faces of the edges of $c$ fixed. The notation with the sum and products in eq.~(\ref{eq:transitionamplitudequantumgroup-5}) 
is to indicate that one needs to use all possible configurations allowed.

\subsection{Fibonacci Fusion Hilbert space}

We now consider the topological data of $SU(2)^5_q$.
The  admissible spin representations are
$j=0,\frac{1}{2},1,\frac{3}{2}$, and their quantum dimension are given by 
\begin{equation}
d_{j}^{q}=sin\left(\frac{\pi(2j+1)}{5}\right)/sin\left(\frac{\pi}{5}\right).
\label{eq:quantumdimension}
\end{equation}
The composition of representations 
follows the following fusion rules:
\begin{eqnarray}
0\otimes j & =j\nonumber \\
\frac{3}{2}\otimes j & =\frac{3}{2}-j\nonumber \\
\frac{1}{2}\otimes\frac{1}{2} & =0\oplus1\nonumber \\
\frac{1}{2}\otimes1 & =\frac{1}{2}\oplus\frac{3}{2}\nonumber \\
1\otimes1 & =0\oplus1.\label{eq:fusionrulessu23}
\end{eqnarray}
Remember that a segment of the 
triangulation is dual to a face on the dual 2-complex where there is associated a representation of $SU(2)^5_q$.
So more fundamentally there are 2 states on the segments, one with quantum dimension $1$
and one with quantum dimension $\phi$, and we have a notion of a 2-dimensional 
fusion Hilbert space \cite{ZhenghanWang,Pachos} 
as the superposition 
of these 2 states. The space $H_{j_{q1}}\times H_{j_{q2}}\times H_{j_{q3}}$
is in this way 3-dimensional and is related to the triangular faces. 
The tetrahedron $H_{j_{q1}}\times H_{j_{q2}}\times H_{j_{q3}}\times H_{j_{q4}}$
is 5 dimensional. The dimension grows with increasing number of representations, following the well known
Fibonacci sequence, which gives the name of Fibonacci anyons for the representations of the $SU(2)^5_q$ symmetry
in context of topological phases of matter and quantum computing field of researches.
In this language the  $A_{v}=W_{\triangle_{1}}^{5}(j_{lv})$
building block amplitude is just the known F symbol that appears in those anyonic models \cite{ZhenghanWang,Pachos}. 
The partition function is about gluing these $A_{v}$
amplitudes, which makes sense from the point of view of quantum computation.
An important result in the field of quantum computation is that a 
higher dimensional Hilbert space can be decomposed into a network of qubits and 2-level quantum gates \cite{nielsen_chuang},
and that with at least 3 Fibonacci anyon representations, one can implement the qubits, where the braids of these anyons can 
approximate any 2-level quantum gate, making Fibonacci anyons a sufficient substrate for universal quantum computation \cite{ZhenghanWang}.
So $Z_{q}(nj)$ of $n$ ``anyon states'', for large $n$, is a topological quantum
liquid or a topological quantum computer but described by the network of the  building
block amplitudes $A_{v}=W_{\triangle_{1}}^{5}(j_{lv})$.
We note that the dual 2-complex of the quantum tetrahedron is made of one vertex in the bulk
and 4 nodes on the triangular boundary faces of the triangulation. So in fact the quantum tetrahedron
carries 4 anyonic charges, the 4-tensor product of the $H_{j_{qi}}$ above, implemented by $A_v$.
The extension of the topological data from  $SU(2)^5_q$ to $SU(N)^r_q$ is presented in the
next section.


\section{Fermionic Cycles and Gauge Symmetry Unification}
\label{sec:unification}

The natural symmetry for the quantum tetrahedron, as discussed in Section~\ref{qt},
is the spatial rotations $SO(3)$ in 3D, implemented by the covering group $SU(2)$, 
and hence its fifth-root-of-unity deformation.
But the generalization
of $SU(2)=A_{1}$ in the Weyl-Cartan or Chevalley basis to the higher dimensional
simple Lie algebras, eq.~(\ref{eq:SU3algebrarelation}), indicates that we can incorporate
the so called internal symmetry or charge space in the same way. The
simplest way is to consider the symmetry $SU(2)\times G$, with $G$
one of the higher dimensional simple Lie algebras \cite{Bianchi:2010bn}. Because of the
importance of $A_{1}$ the natural first candidate is $G=A_{2}=SU(3)$.
Following this, we will also define an extension of the topological data from  $SU(2)^5_q$ to $SU(N)^r_q$.
To get onto multi-dimensional Fibonacci-like anyonic representations and fulfill
the quantizing postulate eq.~(\ref{eq:commutationJplusminos-1-1}), 
we will further restrict $r$ to be a multiple of $5$ larger than $N$.
Examples useful for gauge unification are given by: $SU(3)^5_q$, $SU(4)^5_q$, $SU(5)^{10}_q$, $SU(8)^{10}_q$ and $SU(9)^{10}_q$.
This allows us to address the higher dimensional $A_{4}$,
whose Coxeter-Dynkin diagram splits into two copies of $A_{2}$, and $E_{8}$,
the largest exceptional Lie algebra, which
has four $A_{2}$ building blocks as shown in Figure~\ref{ADE8_Dynkin_Diagrams}. 
As is well known, $A_{4}=SU(5)$ and $E_{8}$
are very important for unification physics. We will also discuss the exceptional Lie 
algebra $G_2$, which contains $SU(3)$ and has a fifth-root deformation with Fibonacci
fusion rules.


\subsection{Gauge Unification Physics Review}
\label{gaugeunireview}
Let us provide a short review of the Lie algebra unification physics program closely following the references \cite{Georgi,Koca3,Koca4}.
The well known representation theory of $SU(2)$ can be extended to higher dimensional Lie groups and algebras.
The structure of a simple Lie algebra is described by its root and weight systems. 
The classification of simple Lie algebras is an extensive subject on which we recommend the references \cite{Georgi,Zee,gilmore}.
Here we collect and review just the elements necessary for our discussion.
For example we have that the root system $A_n$ describes $SU(n+1)$, while $D_n$, with $n\geq3$, describes the Lie algebra of $SO(2n)$.
The exceptional Lie algebras $G_2$, $F_4$, $E_6$, $E_7$ and $E_8$ have their root systems with the same respective names.
For our discussion we will focus on $A_n$, in particular $A_2$, and $E_8$, 
whose importance for coupling the charge space with spin foam was pointed out in
\cite{Aschheim2}, and which contains the other exceptional Lie algebras as well the $A_n$ for lower $n$.

\begin{figure}[!h]
	\centering{}
	\includegraphics[scale=0.30]{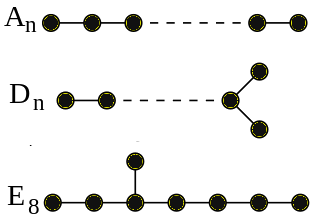}
	\caption{Coxeter-Dynkin diagrams of $A_n$, $D_n$ and $E_8$.}
	\label{ADE8_Dynkin_Diagrams}
\end{figure}

Let us consider a Euclidean space $V$ of dimension $n$. One can introduce an orthonormal
basis, $l_{i}$ $(i=1,2,...,n)$, in $V$ so that the Euclidean scalar
product $(u,v)$ of $u=\sum_{i}\lambda_{i}l_{i}$ and $v=\sum_{i}\mu_{i}l_{i}$
is $(u,v)=\sum_{i}\lambda_{i}\mu_{i}$.
The Coxeter-Dynkin diagrams of $A_n$, $D_n$ and $E_8$ are shown in Figure~\ref{ADE8_Dynkin_Diagrams}. 
The nodes on the Coxeter-Dynkin diagram represent the simple roots
$\alpha_i$, $(i=1,2,...,n)$, of the associated Lie algebra of rank $n$\footnote{An $N$ dimensional Lie algebra is a vector space 
which contains an $n$ dimensional subspace, the so called \textbf{Cartan subalgebra}, spanned by a maximal set of $n$ inter-commuting generators, $H_a$,
$\big[H_{a},H_{b}\big]=0\qquad\forall\quad1\le a,b\le n$.
} and two dots linked by an edge correspond to two roots whose scalar product is $-1$. 
The other pairs of nodes, which are not connected, correspond to vectors that are orthogonal.
The norms are given by $(\alpha_i,\alpha_i)=2$ and the Cartan matrix elements are
\begin{equation}
C_{ij}=\frac{2(\alpha_{i},\alpha_{j})}{(\alpha_{j},\alpha_{j})}.\label{eq:cartanmatrix}
\end{equation}
Complementary to the simple roots, there are the fundamental weight vectors $\omega_{i}$, 
which are defined by the relation with the simple roots
$(\omega_{i},\alpha_{j})=\delta_{ij}$ and are related to each other by 
$(\omega_{i},\omega_{j})=(C^{-1})_{ij}$. 
The root lattice $\Lambda$
is the set of vectors $p=\sum_i^n b_{i}\alpha_{i}$,
$b_{i}\in \mathbb{Z}$. Similarly the weight lattice $\Lambda^{*}$
is the set of vectors $q=\sum_i^n p_{i}\omega_{i}$,
$p_{i}\in\mathbb{Z}$. The reflection
generator with respect to the hyperplane orthogonal to the simple
root $\alpha_{i}$ is $r_{i}$, $(i=1,...,n)$, which operates on an arbitrary vector $\lambda$
as
\begin{equation}
r_{i}\lambda=\lambda-\frac{2(\lambda,\alpha_{i})}{(\alpha_{i},\alpha_{i})}\alpha_{i}.\label{eq:Weylreflection}
\end{equation}
It transforms a fundamental weight vector as $r_{i}\omega_{j}=\omega_{j}-\alpha_{i}\delta_{ij}$.
These generators form the Coxeter reflection group (Weyl
group) acting on the root system $G$ defined by the presentation $W(G)=\langle r_{1},...,r_{n}|(r_{i}r_{j})^{m_{ij}}=1\rangle$.
When they act on one of the simple roots of a simple
Lie algebra they generate the root system $W(G)\alpha_{i}$. 

\begin{figure}[!h]
	\centering{}
	\includegraphics[scale=0.30]{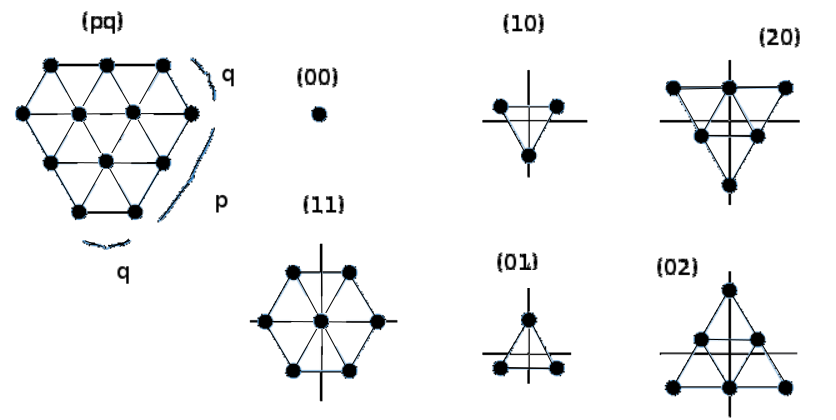}
	\caption{$A_2$ ``polytopes''.}
	\label{A2polytopes}
\end{figure}

The concept of polytopes is important to our discussion. The root
polytope is a convex polytope whose vertices are vectors of the root
system.
The highest weight vector for an irreducible representation of the
Lie algebra is defined as the weight vector $\omega_{rep}=(p_{1},...,p_{n})$,
where $p_{i}$ are positive integers called Dynkin labels. 
The highest weight of any irreducible representation decomposes on fundamental weights, 
and its components are its Dynkin labels
\begin{equation}
\omega_{rep}=\sum_{i}p_{i}\omega_{i}.\label{eq:highestweightdecomposition}
\end{equation}
The highest weight vector has an associated polytope,
which is the convex polytope
possessing the symmetry of the Coxeter-Weyl group as the orbit of
this highest weight vector $W(G)(p_{1},...,p_{n})=(p_{1},...,p_{n})_{g}$.
With this notation the root polytope of $W(A_{n})$ is given, in a simplified form,
by $(10...01)_{A_{n}}$. For $A_{2}$, for example, $(11)$ describes the 
root polytope, the hexagon with 2 points in the middle in Figure~\ref{A2polytopes},
and each point corresponds to one of the 8 states of the adjoint representation, so $(11)$ can
refer also to the whole adjoint representation.
The fundamental simplex of the lattice
$A_{n}$ is a convex polytope with $n+1$ vertices given by $\omega_{1},...,\omega_{n}$
and the origin $(0)$.
The Voronoi polytope $V(0)$ centered at the origin of the lattice
$\Lambda$ is the set of points $V(0)=\left\{ x\in\mathbb{R}^{n},\forall p,(x,p)\leq\frac{1}{2}(p,p)\right\} $.
Let $v\in\mathbb{R}^{n}$ be a vertex of an arbitrary Voronoi polytope
$V(p)$. The convex hull of the lattice points closest to $v$ is
called the Delone polytope containing $v$. Vertices of the Voronoi
cell $V(0)$ of the root lattices of the $A_{n}$ series consist
of the vertices of the Delone cells centered at the origin.

Let us show the analogue of these polytopes in lower dimension for $A_1=SU(2)$. 
First write the eigenvalues in integer form, $p=2j$ and then:
\begin{itemize}
\item
$p=1$ is the highest weight of the fundamental representation with one
reflection that sends it to the state $p=-1$. The edge from the $p=-1$ to $p=1$
is the Voronoi polytope, written $(1)$ in the notation above.
\item 
With $p=2$, operating with the $W(A1)$ one gets the other 2 states of the adjoint 
representation, $p=0$ and $p=-2$. The edge from $p=-2$ to $p=2$ forms the root polytope which is the dual
of the Voronoi polytope, and is represented as $(2)$, which also indicates the adjoint representation.
\item
Note that the building block is the Delone polytope which is an edge of lenght 1.
\end{itemize}

For $A_2$ we present the representations of interest in Figure~\ref{A2polytopes}, 
and note that the Voronoi cell is the hexagon made of the 2 triangles (10) and (01), 
while the Delone cell's building blocks are the triangles whose centers are the vertices of the 
Voronoi cell. Each one is equivalent to one of the triangles (10) and (01) $-$ and 
six of them make the hexagon root polytope.

It is not necessary to introduce the polytopes of the exceptional Lie algebras because we will 
work with $G_2$ and $E_8$ only as a container for the interaction between its $A_2$ subalgebras,
which will be our main objects.

 To present a general representation of the roots and the weights
of $A_{n}$ we add one vector to our orthonormal set of
vectors, $l_{i}$ $(i=1,2,...,n+1)$, $(l_{i},l_{j})=\delta_{ij}$ . 
We define the simple roots as linear combinations of orthonormal
vectors $\alpha_{i}=l_{i}-l_{i+1}$ , $(i=1,2,...,n)$. The group
generators $r_{i}$ permute the set of orthonormal vectors as $r_{i}:l_{i}\leftrightarrow l_{i+1}$. 
When we define the vectors in terms of their components in the $n+1$
dimensional Euclidean space the simple roots and the fundamental weights
read 
\begin{eqnarray}
\alpha_{1} & = & (1,-1,0,...,0);...;\alpha_{n}=(0,0,...,1,-1);\nonumber \\
\omega_{1} & =\frac{1}{n+1} & (n,-1,...,-1);\omega_{2}=\frac{1}{n+1}(n-1,n-1,-2,...,-2);\nonumber \\
...; & \omega_{n}= & \frac{1}{n+1}(1,1,...,1,-n).\label{eq:rootsweightAn}
\end{eqnarray}
For $SU(3)$ we have the simple roots and weights
\begin{eqnarray}
\alpha_{1}= & (1,-1,0),\nonumber \\
\alpha_{2}= & (0,1,-1),\nonumber \\
\omega_{1}= & (2/3,-1/3,-1/3),\nonumber \\
\omega_{2}= & (1/3,1/3,-2/3).\label{eq:simplerootweighta2}
\end{eqnarray}
They can be used to compute the scalar products on the amplitudes
discussed below.


The derivation of the Lie algebra from its root system is standard.
Let us consider the generalization of eq.~(\ref{eq:commutationJplusminos-1}) for $SU(2)$ to $SU(3)$.
To each of the 2 simple roots $\alpha_{i}$ we associate three generators $h_{i}$,
$L_{i}^{+}$ and $L_{i}^{-}$. These generate the $SU(3)$ algebra
subject to the relations 
\begin{eqnarray}
\left[h_{i},h_{j}\right] & =0\nonumber \\
\left[h_{i},L_{j}^{\pm}\right] & =\pm C_{ij}L_{j}^{\pm}\nonumber \\
\left[L_{i}^{+},L_{j}^{-}\right] & =\delta_{ij}h_{j},\nonumber \\
\left[L_{i}^{\pm},\left[L_{i}^{\pm},L_{j}^{\pm}\right]\right] & =0, & i\neq j\label{eq:SU3algebrarelation}
\end{eqnarray}
where, from eq.~(\ref{eq:cartanmatrix}), $C_{ij}=-1+3\delta_{ij}$. 
The Cartan-Weyl basis, or in this situation the Chevalley basis, allows us to write the large dimensional
Lie algebra in terms of the relationship between its building block $SU(2)$
subalgebras.

\subsection{Spin foam at the fifth root of unity with fermionic cycles and gauge symmetry}

Following \cite{You_jin_1992,Ardonne}, the deformation of the $SU(3)$ algebra ($SU(N)$ in general) eq.~(\ref{eq:SU3algebrarelation}) at the fifth root of unity
is given by:
\begin{eqnarray}
\left[h_{i},h_{j}\right] & =0\nonumber \\
\left[h_{i},L_{j}^{\pm}\right] & =\pm C_{ij}L_{j}^{\pm}\nonumber \\
\left[L_{i}^{+},L_{j}^{-}\right] & =\delta_{ij}\left[2h_{i}\right]_{q},\nonumber \\
\left(L_{i}^{\pm}\right)^{2}L_{j}^{\pm}-\left[2\right]_{q}L_{i}^{\pm}L_{j}^{\pm}L_{i}^{\pm}+L_{j}^{\pm}\left(L_{i}^{\pm}\right)^{2} & =0, & i\neq j\label{eq:SU3algebrarelationquantum5throot}
\end{eqnarray}
which reduces to eq.~(\ref{eq:SU3algebrarelation}) when $q$ goes to 1. From eq.~(\ref{eq:qnumber}), $\left[2\right]_{q}=\phi$ in our case.

$SU(3)$ has two Cartan generators and as in the case of $SU(2)$ we label its representations by
the eigenvalues of the Cartan generators. So the representations are
labeled by 2 Dynkin labels $g=(p_1p_2)$ in the notation introduced above for their associated ``polytopes'', 
or in a more compact form by
its classical dimension 1, 3, $\bar{3}$, 6, $\bar{6}$, 8 and so
on. For $SU(3)_{q}^{5}$ the fusion rules are limited to the lower
dimensional representations $1=(00)$, $3=(10)$, $\bar{3}=(01)$,
$6=(20)$, $\bar{6}=(02)$ and $8=(11)$.
The fusion rules are given
by $(00)\otimes (p_1p_2)=(p_1p_2)$ and Table~\ref{tablefusionrulessu3}.
\begin{table}
\begin{centering}

\begin{tabular}{|c|c|c|c|c|c|}
\hline 
$\otimes$ & (10) & (01) & (20) & (02) & (11)\tabularnewline
\hline 
\hline 
(10) & $(01)\oplus(20)$ &  &  &  & \tabularnewline
\hline 
(01) & $(00)\oplus(11)$ & $(10)\oplus(02)$ &  &  & \tabularnewline
\hline 
(20) & (11) & (10) & (02) &  & \tabularnewline
\hline 
(02) & (01) & (11) & (00) & (20) & \tabularnewline
\hline 
(11) & $(01)\oplus(02)$ & $(01)\oplus(20)$ & (01) & (10) & $(00)\oplus(11)$\tabularnewline
\hline 
\end{tabular}
\captionof{table}{Fusion rules of $SU(3)_{q}^{5}$ \label{tablefusionrulessu3}}

\par\end{centering}

\end{table}

To compute for general $SU(N)_q$ the quantum dimensions $d_{q}(g)$, which will be the 
edge amplitudes, 
the spin number $j$ of equation (\ref{eq:quantumdimension}), for a specific representation, 
is extended to a Dynkin label
 $(p_{1},...,p_{N-1})$ describing the highest 
 weight of the representation. 
 For general $N$, $d_q(p_{1},...,p_{N-1})$
will be computed  by the following algorithm: a
Young diagram is drawn for the Dynkin label as a set of $N$ left-justified
rows of $p_{1}$, $p_{2}$, ... $p_{N-1}$ cells, from top to bottom.
If $p_{N-1}>0$ there exists a minimal $M$ index, $0<M<N$ such that
$p_{M+1}=0$, and we draw only $M$ rows, otherwise we set $M=N-1$. 
Accumulating
its value from the right to the left, we create a partition numbering
the length of successive rows in the Young diagram. The Young diagram
can be filled by numbers beginning at the top-left by an integer between
1 and $N$, non increasing from left to right and strictly decreasing
from top to bottom, which enable to compute the number of elements
corresponding to different Young tableaux. This is the dimension of
the representation. It can be expressed by the Weyl dimension formula
as the ratio: the numerator is the product of the number in each box of a tableau
beginning with $N$ at the top-left, strictly increasing to the left
and strictly decreasing to the bottom, while the denominator is the product of the
hook number of each box, where the hook number of any box is the number
of boxes met by a hook beginning at the right of the row, tracing
horizontally to the box, then tracing vertically to the bottom. Because
the numerator will vanish if there are more than $N-1$ rows, the
number of rows is limited. For the quantum deformation, the same Weyl
formula is applied but the integers in each box are now quantum
integers defined by eq.~(\ref{eq:qnumber}), where $[r]_{q}=0$.
Therefore the Young diagram cannot have more than $r-N$ columns,
which sets a cut-off for the number of representations. For $N=3$
the quantum dimension is given by 
\begin{equation}
d_{q}(p_{1},p_{2})=[p_{1}+1]_{q}[p_{2}+1]_{q}[p_{1}+p_{2}+2]_{q}/[2]_{q}.\label{eq:quantumdimensionsu3}
\end{equation}
 The quantum dimension $d_{q}(g)$ for $g=(00), (20), (02)$ is $d_{q}(g)=1$ and for $g=(10), (01), (11)$
 is $d_{q}(g)=\phi$.
 For general $N$ the quantum dimension is gives by \cite{Zub15}
\begin{equation}
d_q(p_1,p_2, ..,p_{N-1})=\frac{[N-1]_q!}{\prod_i [b_i]_q!}\prod_{0<i<j<N}[b_i-b_j]_q,
\textit{ ~with~}
b_i = \sum_{j=i}^{N-1}p_i,
\label{eq:nquantumdimension}
\end{equation}
and where $[m]_{q}!=\prod_{n=1}^{m}[n]_{q}$.

Alternatively,  we can use
the general formula from which
eq.~(\ref{eq:quantumdimension}) is derived, a deformation of the Weyl dimension formula to
\begin{equation}
d_q(g)=\prod_{\alpha}\frac{\left[(\omega_{g}+\rho,\alpha)\right]_{q}}{\left[(\rho,\alpha)\right]_{q}},\label{eq:Weylformula}
\end{equation}
where $\omega_{g}=(p_{1},...,p_{N-1})$, the highest weight describing the irreducible representation,
is given by eq.~(\ref{eq:highestweightdecomposition}).
The vector $\rho$ is half the sum of positive roots $\rho=\frac{1}{2}\sum_{\alpha}\alpha$
and $\alpha$ represent the positive roots. For $SU(2)$ there are only
one positive root $\alpha_{1}=(1,-1)$ and the fundamental weight
is given by $\omega_{1}=(1/2,-1/2)$, so $\rho=(1/2,-1/2)$ and $\omega_{g}=p_{1}\omega_{1}=2j\omega_{1}$
and eq.~(\ref{eq:Weylformula}) reduces to eq.~(\ref{eq:quantumdimension}).
For $N=3$, from eq.~(\ref{eq:simplerootweighta2}), the third positive root is 
$\alpha_{3}=(1,0,-1)$, $\rho=(1,0,-1)$ and
$\omega_{g}=p_{1}\omega_{1}+p_{2}\omega_{2}$. Thus the 
quantum dimension reduces to eq.~(\ref{eq:quantumdimensionsu3}).

 We now compute the $\{6g\}^{5}$ symbols associated with a tetrahedron edge decoration 
 by representations of the $g=SU(3)_{q}^{5}$ algebra,
but it can be generalized to $g=SU(N)_{q}^{r}$. 
We have adapted the formulas from \cite{Kir88}, replacing
spins by a suitable projection of the highest weight vectors, including
the $1/2$ factor needed for $SU(2)$ and a re-scaling due to dimension.
This projection operator depends on the algebra but not on the level:
\begin{equation}
(w_{1}w_{2}...w_{N-1})'=Pr(N,(w_{1}w_{2}...w_{N-1})):=w.\Omega_{N}\label{eq:projection}
\end{equation}
and we have
\begin{eqnarray}
\Delta'(abc)= & ([(-a+b+c)']![(a-b+c)']![(a+b-c)']!)^{1/2}\nonumber \\
 & \times([(a+b+c)'+1]!)^{-1/2}\label{eq:delta}
\end{eqnarray}
and
\begin{eqnarray}
\{6g\}^{5}(abcdef) & = & \left\{ \begin{array}{ccc}
a & b & e\\
d & c & f
\end{array}\right\} _{q}=\nonumber \\
 & = & \Delta'(abe)\Delta'(acf)\Delta'(ced)\Delta'(dbf)\sum_{z}\frac{\nu}{\delta},\label{eq:symbol}
\end{eqnarray}
where $z$ is an integer vector of ${\mathbb{N}}^{(N-1)}$ inside the polytope
where the expression is not null, and $\nu$ and $\delta$ are
\begin{eqnarray}
\nu & = & (-1)^{z'}[(z+1)']![(z-a-b-e)']![(z-a-c-f)']!\nonumber \\
 &  & \times[(z-b-d-f)']![(z-d-c-e)']!\label{eq:nu}
\end{eqnarray}
\begin{equation}
\delta=[(a+b+c+d-z)']![(a+e+f+d-z)']![(e+b+c+f-z)']!.\label{eq:denominator}
\end{equation}
Finally, the F-symbol ($(F_{c}^{abd})_{ef}$) is given from:
\begin{equation}
\left\{ \begin{array}{ccc}
a & b & e\\
d & c & f
\end{array}\right\} _{q}=(-1)^{(a+b+c+d+e+f)'}([e'][f'])^{-1/2}(F_{c}^{abd})_{ef},\label{eq:Fsymbol}
\end{equation}
where $[e']$ is the quantum dimension of the representation of highest
weight $e$. The fusion matrix ($\left[\begin{array}{cc}
a & b\\
d & c
\end{array}\right]$) is obtained from:
\begin{equation}
(F_{c}^{abd})_{ef}=a_{ef}\left[\begin{array}{cc}
a & b\\
d & c
\end{array}\right].\label{eq:fusionmatrix}
\end{equation}

For $N=3$, we can use either the $F$-symbols from \cite{Ardonne}
or the fusion matrix from \cite{Nav13}. The following are two examples
showing the coherence of our formulas (\ref{eq:Fsymbol}) and (\ref{eq:fusionmatrix})
with some results given in \cite{Kir88,Ardonne,Nav13}:
\begin{eqnarray}
\left\{ \begin{array}{ccc}
01 & 10 & 00\\
01 & 01 & 00
\end{array}\right\} _{q} & = & (-1)^{(13)'}([00'][00'])^{-1/2}(F_{01}^{01~10~01})_{00~00}\nonumber \\
 & = & (-1)^{0}(1)^{-1/2}(F_{\bar{3}}^{\bar{3}~3~\bar{3}})_{1~1}=(F_{3}^{3~\bar{3}~3})_{1~1}\nonumber \\
 & = & a_{00~00}\left[\begin{array}{cc}
01 & 10\\
01 & 01
\end{array}\right]=[3]_q^{-1}\label{eq:examplesu36jq1}
\end{eqnarray}
and
\begin{eqnarray}
\left\{ \begin{array}{ccc}
10 & 01 & 00\\
01 & 01 & 10
\end{array}\right\} _{q} & = & (-1)^{(23)'}([00'][10'])^{-1/2}(F_{01}^{10~01~01})_{00~10}\nonumber \\
 & = & (-1)^{1}(1)^{-1/2}(F_{\bar{3}}^{3~\bar{3}~\bar{3}})_{1~1}=-(F_{3}^{\bar{3}~3~3})_{1~3}\nonumber \\
 & = & -a_{00~10}\left[\begin{array}{cc}
10 & 01\\
01 & 01
\end{array}\right]=[3]_q^{-1/2}.\label{eq:examplesu36jq1-1}
\end{eqnarray}

A complete algorithm to compute $SU(4)$ Clebsch-Gordan coefficients
is available in \cite{Kuh08}; it can be extended to compute the 
{6j}-symbols of $SU(4)$ for all representations. The algorithm may be extended
to higher dimensions and deformed by a procedure replacing factorials
by quantum factorials and using projected weights.
The Fibonacci anyons in $g=SU(4)_{q}^{5}$ will be restricted to only 4 representations,
the fundamental $(000)$, the symmetric $(100)$, its conjugate antisymmetric
$(001)$ and the antisymmetric $(010)$. Their respective quantum dimensions
are $[1]_{q}$, $[1]_{q}$, $[4]_{q}$ and $[4]_{q}[3]_{q}$/$[2]_{q}$. 
For $r=5$ these quantum numbers are all equal to 1, therefore this is adapted
to decorate a graph with only one edge length, while $SU(2)_{q}^{5}$
and $SU(3)_{q}^{5}$ decorate graphs with two edge lengths. The
cut-off in the number of representations comes from the fact that
$r-N$ is small. The next representations which would be of interest for a
higher $r$ are the symmetric $(200)$, cut-off at $r=5$, and the symmetric
$(300)$, cut-off at $r=6$. The Young tableau of $(200)$ is a row with two
boxes. In the dimension formula, the numerator is the product of the
two boxes, where the leftmost has the quantum integer $[N]$ and each
one at its right increases by one, so the right one is
$[N+1]=[r]=0$, which cancels this representation.

There is no representation in $SU(5)_{q}^{5}$, so we study $SU(5)_{q}^{10}$
which is quite rich. When $N$ increases to meet $r-1$, the structure
simplifies, and there are only 9 antisymmetric representations in
$SU(9)_{q}^{10}$: $(0^8),(1~ 0^7),(0^1~1~0^6),(0^2~1~0^5)... (0^6~1~0^1)$ and $(0^7~1)$. 
They will be the bricks for the branching of
$E_{8q}^{10}$ to $A_{8q}^{10}$, which is a subject of our ongoing work.

The observables of interest
can be calculated from the generalization of eq.~(\ref{eq:transitionamplitudequantumgroup-5})
to include fermions, which are defined as closed oriented cycles on the dual 2-complex $\bigtriangleup^{*}$, 
with $SU(3)^5_q$ representations on its edges. These fixed states can be interpreted as observables \cite{Garcia-Islas}. 
The observable of interest is
\begin{eqnarray}
O_{\triangle_{4^{l}}}(g_{c},j_{lc}) & = & \mathcal{N}_{q}\sum_{\{c\}}\sum_{j_{f}|j_{lc}}\prod_{f}(-1)^{j_{f}}d_{q}(j_{f})\prod_{v}W_{\triangle_{1}}^{5}(j_{lv})\nonumber \\
 &  & \prod_{c}\left(\prod_{e\in c}(-1)^{o}d_{q}(g_{e})\right)\{6g\}^{5}\label{eq:transitionamplitudequantumgroup-5su3}
\end{eqnarray}
where $o$ gives a sign due to the match or mismatche of the orientations of  $\bigtriangleup^{*}$ and the cycles $\{c\}$.
In this sense the cycles are fermionic $-$ there cannot be more than one cycle on one edge with the same orientation
relative to the orientation of $\bigtriangleup^{*}$. 
The maximum is two cycles on one edge with opposite orientation, which constrains the possible cycles.
Each edge on a cycle $c$ has one amplitude given by the quantum dimension $d_{q}(g_{e})$ of the $SU(3)^5_q$ 
representation $g$  on it.
The $\triangle_{4^l}$ is a triangulation at level $l$, which we can explain with examples:
\begin{itemize}
\item
 The level $l=0$ triangulation is just one tetrahedron in $\triangle$ 
 and one vertex inside the boundary tetrahedron. 
 \item
 At the level $l=1$ triangulation we start with one tetrahedron and connect its
center with its 4 points, making four new tetrahedrons. Then we take the dual
2-complex in the usual way by inserting a vertex inside each tetrahedron and connecting them. 
(See Figure~\ref{triangduallevel1}). 
\item
At the level $l=2$  triangulation, our space is divided into 16 tetrahedrons.
The dual 2-complex in the bulk has 16 vertices at their centers. (See Figure~\ref{qg5_cycles}). 
And so on.
\end{itemize}

The observables $O_{\triangle_{4^{l}}}(g_{c},j_{lc})$ are functions of the 
$SU(2)^5_q$ $j_{lc}$ representations on the boundary faces of $\bigtriangleup^{*}$ and 
$SU(3)^5_q$ $g_{c}$ on the edges of the closed cycles.
To correctly assign the $\{6g\}^{5}$ on the network, we need to consistently choose the cycles and 
we need to give some consideration to how to constraint our cycles of interest.
Let us discuss this level by level.

\subsubsection*{The level $l=0$ observables:}

The level $l=0$ has no bulk edges and so there are only pure
 topological (gravitational) observables, eq.~(\ref{eq:transitionamplitudedelta1-5}).

\subsubsection*{The level $l=1$ observables and its $\{6g\}^{5}$:}

We note that the usual observables at $l=0$ above, $W_{\triangle_{1}}^{5}(j_{lv})$,
are functions of
the six representations on the six boundary faces of the 2-complex, or 
equivalently on the segments (edges of the tetrahedron) of
the triangulation. We can then associate a ``polytope'' of $SU(2)$ to these
representations; they are the edges of the tetrahedron, and the building
block is the Delone polytope, the edge of $p=1$ of the (1) representation. 
The only representation that does not have this interpretation is the scalar (0),
which can work as ``dummy'' indices in the $W_{\triangle_{1}}^{5}(j_{lv})$.
This suggests that the observables of
interest for $SU(3)_{q}^{5}$ are the ones that have a ``polytope'' interpretation
in $SU(3)^5_q$ so that the $\{6g\}^{5}$ would be a function of these ``polytopes''.
In particular, these would be the Delone polytopes, (10) or (01), and, when necessary, a ``dummy'' index with 
the (00) representation 
to complete the $\{6g\}^{5}$ input representations. 

Immediately we note that, consistently, each edge of a cycle on $\bigtriangleup^{*}$
is dual to a triangle of the $\bigtriangleup$ and so it can have a representation $g=(10),(01)$ or $g=(00)$ 
sitting on it. 
From Figure~\ref{triangduallevel1} let us label the representations: 
\begin{itemize}
\item
$SU(2)_{q}^{5}$ on the faces of $\bigtriangleup^{*}$, using
the numbers labeling the vertices: 
$j_{1}=j_{12}$, $j_{2}=j_{13}$, $j_{3}=j_{14}$, $j_{4}=j_{23}$,
$j_{5}=j_{24}$, $j_{6}=j_{34}$, $j_{7}=j_{123}$, $j_{8}=j_{124}$,
$j_{9}=j_{134}$, $j_{10}=j_{234}$.
\item
And for $SU(3)_q^5$ on the edges from the 4 internal vertices:
$g_{1}=g_{12}$, $g_{2}=g_{13}$, $g_{3}=g_{14}$, $g_{4}=g_{23}$,
$g_{5}=g_{24}$, $g_{6}=g_{34}$.
\end{itemize}
So, for the observable $O_{\triangle_{4}}(g_{1...6},j_{1...6})$ 
the $\{6g\}^{5}$ is naturally associated with the bulk tetrahedron with vertices 1,2,3,4,
coupling six $g$ representations on its six edges. So the set of allowed cycles $\{c\}$ are the ones
covering this tetrahedron, and this can be done with 4 
triangular cycles. Note that any configuration with all (01) or (10),
e.g. $g_{1...6}=((10),(01),(10),(10),(10),(01))$, gives $\{6g\}^{5}=0$.
But by inserting the representation (00) we can have non-zero $\{6g\}^{5}$ and the observables can be computed,
as for example for $g_{1...6}=((01),(10),(01),(01),(00),(00))$, which gives,
from eq.(\ref{eq:examplesu36jq1}), $\{6g\}^{5}=\phi^{-1}$.
In this case, for the amplitude $\{6g\}^{5}$ we are using the fundamental (10) and anti-fundamental
(01) representations which compose the Voronoi polytope of $SU(3)$.

One example of interesting an observable, which can be interpreted as a
coupling interaction of the $SU(2)_{q}^{5}$ and $SU(3)_{q}^{5}$ representations,
is defined in the following way. First we choose an orientation on $\bigtriangleup^{*}$
to be given by one incoming edge from the boundary to vertex 4, then to vertices 1, 2 and 3. 
From 3, the oriented edges go to 1, 2 and from 1, 2 they go out (see Figure~\ref{triangduallevel1}). The
cycle $c_{1}$ is 123, $c_{2}$ is 142, $c_{3}$ is 134 and $c_{4}$
is 243. There is only one other option of cycles, which is the one
with the inversion of orientation of all 4 cycles. 
For $g_{1...6}=((10),(01),(01),(01),(00),(10))$
fixed, from eq.(\ref{eq:examplesu36jq1-1}), $\{6g\}^{5}=\phi^{-1/2}$, 
imposing consistency of both representations on $\bigtriangleup$
(where, for example, a $g_1=(01)$ fixes $j_1=1/2$ ($p=1$, $j=2p$))
 leads to $j_{1...10}=(\frac{1}{2},\frac{1}{2},\frac{1}{2},0,0,0,0,\frac{1}{2},0,\frac{1}{2})$.
Thus, working with the unnormalized amplitudes, we have
\begin{eqnarray}
O_{\triangle_{4}}(g_{1...6},j_{1...10}) & = & (-1)^{j_{1}}...(-1)^{j_{10}}d_{q}(j_{1})...d_{q}(j_{10})W_{\triangle_{1}}^{5}(j_{v_{1}})...W_{\triangle_{1}}^{5}(j_{v_{4}})\nonumber \\
 &  & \left(-d_{q}(g_{1})d_{q}(g_{2})d_{q}(g_{4})\right)\left(d_{q}(g_{1})d_{q}(g_{3})d_{q}(g_{5})\right)\nonumber \\
 &  & \left(d_{q}(g_{2})d_{q}(g_{6})d_{q}(g_{3})\right)\left(-d_{q}(g_{6})d_{q}(g_{4})d_{q}(g_{5})\right)\nonumber \\
 &  & \{g_{1}...g_{6}\}^{5}.\label{eq:observablelevel4exemple}
\end{eqnarray}
Note that in this example all internal $j$'s are fixed by fixing the $g$'s, and so we could first have 
fixed all the 10 $j$'s, which gives a topological invariant observable of $SU(2)_q^5$ \cite{Garcia-Islas}, and the
$g$'s can be interpreted as emergent properties of the network.

\subsubsection*{The level $l=2$ observables and its $\{6g\}^{5}$:}

\begin{figure}[!h]
	\centering{}
	\includegraphics[scale=0.50]{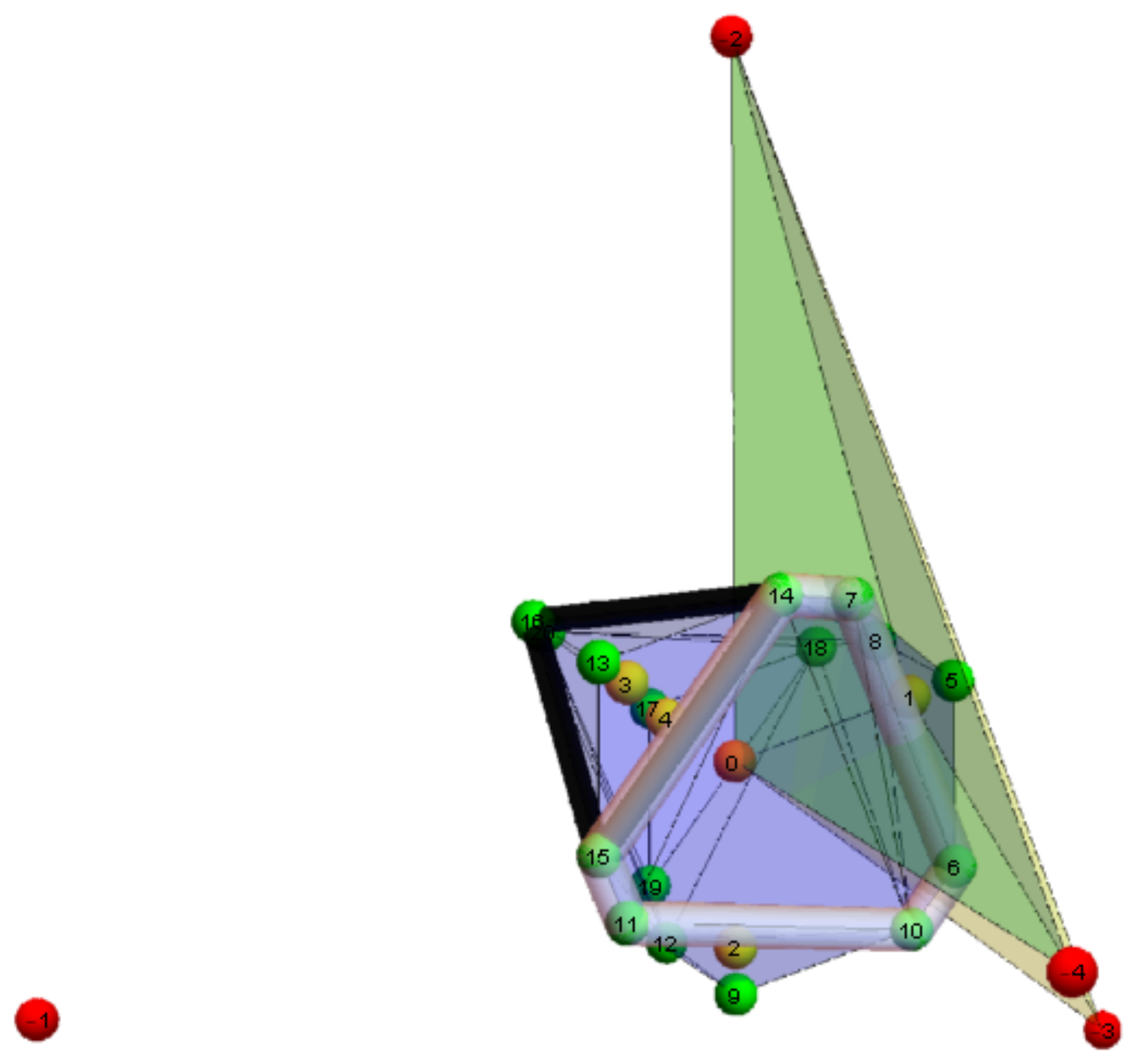}
	\caption{This figure shows the hexagonal cycle $c_1$ in white and the triangular cycle $c_{1.2}$ in black. They share the link l(14-15).}
	\label{qg5_cycles}
\end{figure}

Now we have 4 tetrahedrons similar the ones on $l=1$.
They are connected to form 
4 hexagonal cycles (see Figure~\ref{qg5_cycles}, where internal vertices are labeled with positive integers 
and external points with negative integers):
$c_1$=\{7,14,15,11,10,6\},
$c_2$=\{6,10,12,19,18,8\},
$c_3$=\{8,18,20,16,14,7\},
$c_4$=\{11,15,16,20,19,12\},
where $c_n$ is a hexagonal face (irregular hexagon alternating short and long links), viewed from an outside point $-n$.
Along each of these cycles, the 3 long edges are shared with short cycles completed by a vertex making a dual tetrahedron.
For $c_1$: $c_{1.2}$=\{15,14,16 or 13\}, $c_{1.3}$=\{10,11,12 or 9\}, $c_{1.4}$=\{7,6,8 or 5\}.
The link $c_{1.2}$ is shared between $c_1$, $c_3$ and $c_4$, therefore $c_{1.2}=c_{3.2}=c_{4.2}$.
The link $c_{1.3}$ is shared between $c_1$, $c_2$ and $c_4$, so $c_{1.3}=c_{2.3}=c_{4.3}$.
The link $c_{1.4}$ is shared between $c_1$, $c_2$ and $c_3$, so $c_{1.4}=c_{2.4}=c_{3.4}$.
The last short cycles are $c_{2.1}=c_{3.1}=c_{4.1}=$\{18,19,20 or 17\}.
All four short cycles extended as tetrahedrons cover all the 16 level $l=2$ vertices (centers dual to the $l=2$ tetrahedrons), 
here indexed from 5 to 20.

Emerging $\{6g\}^{5}$ symbols:
\begin{itemize}
\item for a tetrahedron 
$\{6g\}^{5}(c1.2)=\left\{ \begin{array}{ccc}
g_{15-14} & g_{14-16} & g_{16-15}\\
g_{15-13} & g_{14-13} & g_{16-13}
\end{array}\right\} _{q}$;
\item for a hexagon
$\{6g\}^{5}(c1)=\left\{ \begin{array}{ccc}
g_{7-14} & g_{14-15} & g_{15-11}\\
g_{11-10} & g_{10-6} & g_{6-7}
\end{array}\right\} _{q}$.
\end{itemize}
(See Figure~\ref{qg5_cycles}). 

\subsection{Lie algebra polytopes and aperiodic order}

We have the polytope interpretation of our observables where for example the amplitudes $W_{\triangle_{1}}^{5}(j_{lv})$ 
are tetrahedrons dual to the vertices with the amplitudes in the 2-complex.
The edges of these tetrahedrons are $SU(2)^5_q$ representations. And we have that the
$\{6g\}^{5}$ couples together $SU(2)^5_q$ representations sitting on edges of the 2-complex or equivalently on
the triangles of the triangulation.
These Lie algegraic polytopes are few for $SU(2)^5_q$: points and edges with quantum dimension 1 or $\phi$. 
And also few for $SU(3)^5_q$: points, triangles for (10) and (01), a large triangle for (20), (02) and the hexagon for (11). 
(See Figure~\ref{A2polytopes}).
In the classical situation one could try to include higher representations because the Voronoi cell of $A_2$, 
the hexagon made of the 2 triangles (10) and (01) tiles its weights lattice $\Lambda^{*}$.
But here we stay with the lower dimensional ones presented above. The fifth root deformation imposes a
cutt-of on the lattice representations, suggesting that the large observables will be networks of these
lower dimensional representations, in particular the triangles and edges, the Delone cells building blocks of
the root and weight lattices.

We can also interpret the geometry of the amplitudes and observables in terms of the network of the quantum dimension.
Note that what appears in the edges of the dual for transition amplitudes or observables are the amplitudes $d_{q}(j_{f})$
and $d_{q}(g_{e})$, which are the quantum dimensions 1 and $\phi$.
So if we want to build a large observable with more cycles respecting the fusion rules,
the geometric tiling pattern that emerges for the amplitudes with these quantum dimensions
has aperiodic order instead of the periodic root and weight lattices. 
In this case what emerges is a different kind of root system which has no associated Lie algebra, i.e. one of 
the noncrystallographic root systems \cite{ChenMoodyPatera}.

\begin{figure}[!h]
	\centering{}
	\includegraphics[scale=0.30]{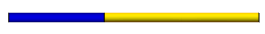}
	\caption{The edges $\{1,0\}$ on the left and $\{0,1\}$ on the right.}
	\label{edgegoldensimplex}
\end{figure}

\begin{figure}[!h]
	\centering{}
	\includegraphics[scale=0.30]{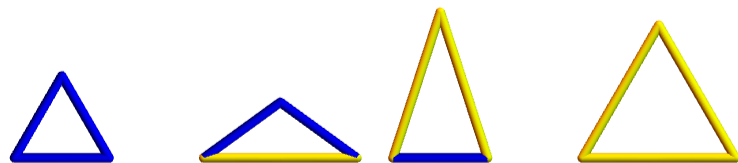}
	\caption{From left to right, $\{3,0\}$, $\{2,1\}$, $\{1,2\}$ and $\{0,3\}$.}
	\label{trianglegoldensimplex}
\end{figure}

\begin{figure}[!h]
	\centering{}
	\includegraphics[scale=0.30]{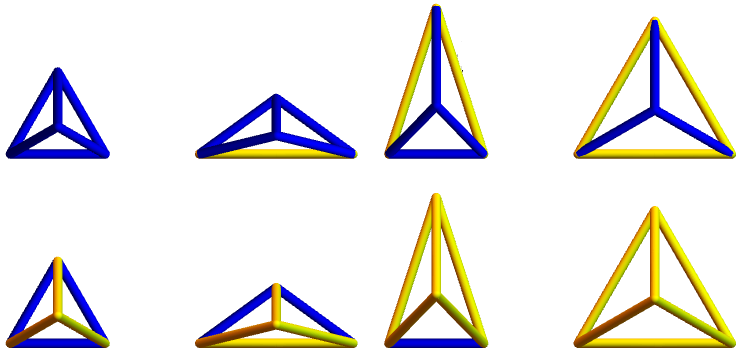}
	\caption{From left to right, in the first line, tetrahedrons $\{6,0\}$, $\{5,1\}$, 
	$\{4,2\}$ and $\{3,3\}$. In the second line tetrahedrons $\{3,3\}$, $\{2,4\}$, 
	$\{1,5\}$ and $\{0,6\}$.}
	\label{tetgoldensimplex}
\end{figure}

To better explain this geometry we introduce the golden simplex, which
we define as a polytope satisfying the following
conditions:
\begin{enumerate}
\item \label{itemenumgolden1}
In a suitable orientation, all coordinates of all vertices of
the ``golden simplex'' take values on a quadratic integer ring $\mathbb{Q}(\sqrt{5})=\mathbb{Z}(\phi)=a+b\phi$, 
 with $a$, $b$ integers.
\item
\label{itemenumgolden2}
Every pair of vertices is separated by an edge whose length is
either $1$ or $\phi$.
\end{enumerate}
Golden edges, triangular faces and tetrahedrons are precisely the geometry of our observables, and
they are consistent with an aperiodic filling of 3D space (which is not possible with regular tetrahedrons) \cite{Kramer82,Kramer91}.
Using the notation $\{m,n\}$ where $m$ counts the number of edges with length $1$ and $n$ the number of edges with length $\phi$, we
have the possibles edge in Figure~\ref{edgegoldensimplex}, the triangular faces in Figure~\ref{trianglegoldensimplex} and the tetrahedrons in Figure~\ref{tetgoldensimplex}.


\subsection{Unification physics from the exceptional Lie algebras}
Let us consider the lowest dimensional exceptional Lie algebra,
$G_2$, at the fifth root of unity $-$ $G^5_{2_q}$. $G_2$, like $SU(3)$, is of rank 2, and is the only exceptional one 
for which a fifth-root quantum deformation is possible. The generalization of eq.~(\ref{eq:SU3algebrarelationquantum5throot}) 
to the exceptional Lie algebras has only 2 equations changed, namely \cite{You_jin_1992,Ardonne}, 
\begin{eqnarray}
\left[L_{i}^{+},L_{j}^{-}\right] & =\delta_{ij}\left[2h_{i}\right]_{q_{i}},\nonumber \\
\sum_{s=0}^{1-C_{ij}}(-1)^{s}\left(\begin{array}{c}
1-C_{ij}\\
s
\end{array}\right)_{q_{i}}\left(L_{i}^{\pm}\right)^{1-C_{ij}-s}L_{j}^{\pm}\left(L_{i}^{\pm}\right)^{s} & =0, & i\neq j\label{eq:G2algebrarelationquantum5throot}
\end{eqnarray}
where $C=\left(\begin{array}{cc}
2 & -3\\
-1 & 2
\end{array}\right)$ and $q_{i}=q^{\frac{1}{t_{i}}}$, with $t_{i}=\frac{2}{(\alpha_{i},\alpha_{i})}$,
which gives for $G^5_{2_q}$, $t_{1}=1$ and $t_{2}=3$. 
The $q_i$ binomials are $\left(\begin{array}{c}
m\\
n
\end{array}\right)_{q_{i}}=\frac{[m]_{q_{i}}!}{[n]_{q_{i}}![m-n]_{q_{i}}!}$, with $[m]_{q_{i}}!=\prod_{n=1}^{m}[n]_{q_{i}}$ and $[n]_{q_{i}}$
given by eq.~(\ref{eq:qnumber}) using $q_{i}$.
There are only
2 representations in the fusion rules, the 1 dimensional $1=(00)$ and
the 7 dimensional, $7=(01)$\footnote{Note that we are using the same notation used for $SU(3)$ in previous sections, 
but the convex polytope possessing the symmetry of the Coxeter-Weyl group, as the orbit of
its highest weight vector, by acting with $W(G_2)$, is a different polytope, a hexagon made of the two 
triangles of $SU(3)$, (10) and (01), and a point in the center, the (00) representation.}. The only non trivial fusion rule is
\begin{eqnarray}
(01)\otimes(01) & =(00)\oplus(01),\label{eq:fusionrulesG2}
\end{eqnarray}
which is the well known Fibonacci (anyon) fusion rule. 
The quantum dimensions, which are the amplitudes on the edges of the $\triangle$*,
are $d^{(00)}_q=1$ and $d^{(01)}_q=\phi$. For the $\{6g\}^{5}$ amplitudes for $G^5_{2_q}$ there are 64 possibles ones 
but only 15 are not null, which constrains the available observables. 
They are given in \cite{Ardonne}. It is therefore easy to compute observables for $G^5_{2_q}$
using eq.~(\ref{eq:transitionamplitudequantumgroup-5su3}) and following the procedure of previous sections.
In the first levels of the triangulation, however, we find no polytope interpretation, 
which would be a hexagon with a point in the center, on the triangulation. 
This leads us to suggest that the exceptional Lie algebras work as a container for 
the lower dimensional Lie algebras, $SU(2)$ and $SU(3)$, which are in fact the gauge symmetries.
Note that the fusion rule of  $\{6g\}^{5}$ contains only the representations that appear in 
the observable eq.~(\ref{eq:observablelevel4exemple}), because the $7$ of $G_2$ contains the 
(10), (01) and (00) of $SU(3)$. This gives a role to $G_2$ as the one that unifies those 
lower dimensional gauge symmetries.

With $G^5_{2_q}$ having only one non-trivial fusion rule, the generalization to higher dimensional Lie algebras
may best be done with the tenth root of unity. Another option is that our observables already capture information
of higher dimension through projection. This can be seen from the Coxeter-Dynkin diagram of $SU(5)$ and $E_8$. 
See for example the magic star projection of $E_8$ to 2D \cite{Piero}.

\section{Conclusions}
\label{sec:conclusion}
In this paper we have presented quantum gravity observables coupling internal gauge symmetry with spacetime symmetry in a spin foam model.
We note that in usual spin foam or state sum models in 3D, there are 2D states, whose evolution is described
by 3D transition amplitudes. Here we presented a Wilson lattice gauge theory perspective where fermionic observables are cycles that
go through the 3D foam. This fixes a notion of 3D states for the 3D theory. Thus the whole discussion of 
transition amplitudes in 3D, which one might ordinarily think of as being
about dynamics, is in fact more about the kinematics of quantum gravity. 
The evolution of these 3D observables is expected to give the true dynamics in the full 4D theory. 
This can give an alternative route to explicit computations of the full spin foam amplitudes, which are a
subject of recent interest \cite{Dona-Sarno,Sarno-Speziale}. It is also well known that the
$\{6j\}_q$ symbols for $SU(2)_q$ give the exponential of the Einstein-Hilbert action in the
limit of large spin quantum numbers, indicating a good semi-classical limit. 
In our case it will be interesting to investigate the limit with large cycles 
through the spin foam as well as the semi classical limit of the symbols from $SU(3)_q$ and $SU(N)_{q}$
in general. 

We also presented a Lie algebra polytope interpretation of the transition amplitudes and observables,
a ``gravitahedra'' \cite{piero2019} for spin foam quantum gravity. 
The representation theory polytopes emerge from the spin foam together with a notion of aperiodic states. 
This work opens up new questions for subsequent research such as questions on 
(1) phase transition and confinement within the anyonic code, 
(2) direct connection with particle physics observables and the notion of extended particle observables, 
(3) better understanding of the aperiodic states and 
(4) the complete formulation with larger Lie algebra amplitudes. 
The fifth root of unity also has limits on the large dimensional algebras which can be addressed $-$ it stops at $SU(5)=A_4$ 
in the $A_n$ series, for example. It will be interesting to investigate the tenth root of unity 
to incorporate larger Lie algebras and quasicrystal projections of large dimensional root and weight lattices.

The transition amplitudes, partition functions and observables discussed in this paper are all finite, 
which is supported by the tetrahedron quantization postulate to impose the fifth root of unity quantization.
The usual manner in which relativistic theory relates mass, energy and 
geometry, together with the conceptual manner in which quantum mechanics
integrates information in the description of 
fundamental physical systems, can be improved by including \textit{computations} 
in a code theoretic framework.  


\end{document}